\documentclass[twocolumn,twoside,slac_two]{revtex4}
\usepackage{graphicx}
\usepackage{fancyhdr}
\begin{document}

\title{VERITAS Results from Deep Exposure on the Distant FSRQ 4C +55.17}

%

\author{Amy Furniss for the VERITAS Collaboration}
\affiliation{University of California Santa Cruz, 1156 High Street, Santa Cruz, CA 95564, USA}
\author{William McConville for the \textit{Fermi}-LAT Collaboration}
\affiliation{NASA Goddard Space Flight Center, Greenbelt, MD 20771, USA}

\begin{abstract}
We present results from a deep VERITAS exposure of the distant ($z=0.89$) flat-spectrum radio quasar (FSRQ) 4C +55.17. The high flux, hard index and steady emission found by
\textit{Fermi} LAT observations make this blazar a promising very-high-energy (VHE; $E\ge$100 GeV) candidate, offering a possibility to clarify the location of FSRQ VHE emission. Non-detection supports the hypothesis that any VHE gamma-rays are produced within and absorbed by the broad-line region while VHE detection would support an emission region outside the broad line region and far
from the base of the jet. This FSRQ additionally provides the possible means, by photon-photon pair production, to constrain the currently available extragalactic background light (EBL) models out to the
groundbreaking redshift of z=0.89. The log-parabolic model that is fitted to the LAT photons allows an extrapolation of the
fit up to VHE while accounting for the gamma-ray absorption by the EBL. The VERITAS upper limit derived from the deep exposure is compared to this extrapolated VHE flux.
\end{abstract}

\maketitle

\thispagestyle{fancy}


\section{VERITAS Instrument}
VERITAS is an array of four 12-meter imaging atmospheric Cherenkov telescopes based near Tucson, Arizona. This instrument is sensitive to very-high-energy (VHE) gamma-ray photons
of energy greater than 100 GeV and has a 3.5 degree field of view. Each telescope has 350 mirrors which reflect light onto a pixelated camera comprised of 499 photomultiplier tubes.  The
instrument has been operating in full array mode since 2007, and has so far detected 23 VHE blazars, having discovered VHE emission from 10 of them.  See \cite{holder} for further details regarding the operation of VERITAS.

\section{The Extragalactic Background Light}
Detection of VHE emitting blazars is made difficult by the pair production between VHE photons and lower energy extragalactic background light
(EBL) photons \citep{2}. The EBL represents the entirety of starlight that has been emitted and reprocessed by dust throughout the history of the
Universe. This interaction reduces the observed flux from extragalactic objects in a distant dependent manner, making the detection of higher
redshift sources 
a challenge.

Measurements of the EBL offer integrated constraints on all of the processes that contribute, e.g. structure formation and galactic evolution \citep{3}. However, due to bright
foregrounds in our solar system and Galaxy, direct measurements of the EBL are difficult. Moreover, a direct measurement would only reflect the current integrated
state, which leaves the challenging task of unraveling the time history of the EBL. This challenge can be overcome when VHE gamma rays are used to probe the EBL through gamma-ray
absorption. The absorption process deforms the intrinsic VHE gamma-ray spectra emitted by extragalactic objects such as blazars. This deformation can be used to
estimate the spectral properties of the EBL out to the redshift of the source \citep{4}.

\section{\textit{Fermi} LAT Shows 4C +55.17 as a Candidate VHE Source}
The flat-spectrum-radio-quasar (FSRQ) 4C +55.17 has been a steady, bright \textit{Fermi} LAT source, as illustrated by the high energy gamma-ray light curve spanning four years (Figure 1) as well as by the various \textit{Fermi} LAT catalog  values reported in Table I.  Data from the first eleven months of operation yielded acceptable fits to a power law, with a relatively bright 1-100 GeV integral flux and a hard index of $\Gamma\sim2$.  As additional statistics were collected, a log parabola was seen to provide an improved fit to the data.  Additionally, this target continues to show a lack of gamma-ray flux variability, unlike other \textit{Fermi} LAT detected FSRQs.

The steady, bright high-energy gamma-ray flux and hard index made this FSRQ a prime candidate for VHE emission. The relatively regular detections of associated photons with energy above 50 GeV (Table II) supports the VHE candidacy and the target was added to the VERITAS observing plan in the spring of 2010.  The log-parabolic fit can be extrapolated into the VHE band to estimate expected flux values above 100 GeV, according to different EBL models as done in Figure 2 for the log-parabolic fit found using the first four years of \textit{Fermi} LAT data.

\begin{table*}[t]
\begin{center}

\begin{tabular}{|l|c|c|c|c|c|c|}
\hline \textbf{}&\textbf{Integral} & \textbf{$\Gamma$}& \textbf{a} & \textbf{b} &
\textbf{Pivot}&\textbf{Reference} \\
\textbf{}& \textbf{1-100 GeV Flux} & \textbf{} &\textbf{} & \textbf{} &
\textbf{Energy}&\textbf{} \\
\textbf{}& \textbf{[ph cm$^{-2}$s$^{-1}$]} & \textbf{} &\textbf{} & \textbf{} &
\textbf{(MeV)}&\textbf{} \\
\hline 0FGL Catalog   &(7.9$\pm$1.0)$\times10^{-9}$ & --& -- & --&--&\cite{5} \\
\hline 1FGL Catalog  &(1.05$\pm$0.06)$\times10^{-8}$ &2.05$\pm$0.03  &-- &-- &--&\cite{6} \\
\hline 2FGL Catalog   &(1.12$\pm$0.04)$\times10^{-8}$& --&  1.83$\pm$0.03&0.07$\pm$0.01 &641.5 &\cite{7} \\
\hline First Four Years&(1.03$\pm$0.03)$\times10^{-8}$ &-- & 1.87$\pm$0.03& 0.06$\pm$0.01  &641.5&this work\\
\hline
\end{tabular}
\caption{\textit{Fermi} LAT observations of 4C +55.17 over four years.  The 0FGL and 1FGL catalogs fit a differential power-law to the data of the form $dN/dE\propto E^{-\Gamma}$.  Additional statistics collected after the first eleven months of operation provide an indication of spectral curvature, showing an improved fit for a log parabola of the form $dN/dE\propto E^{-{\rm a}-{\rm b Log}E}$ in the 2FGL catalog as well as in the analysis which includes the first four years of LAT data.  The relatively high 1-100 GeV integral flux and hard spectrum show this FSRQ to be a promising candidate for VHE emission. }
\label{l2ea4-t1}
\end{center}
\end{table*}

\begin{table}[t]
\begin{center}

\begin{tabular}{|l|c|c|c|}
\hline \textbf{Energy} & \textbf{Time} & \textbf{Separation} &
\textbf{Probability}\\
\textbf{(GeV)} & \textbf{(MJD)} & \textbf{(degree)} &
\textbf{}\\
\hline 141.2& 55115&  0.061&0.985  \\
\hline 135.2&55999 & 0.486 & 0.528 \\
\hline 80.5& 54977& 0.555 & 0.682 \\
\hline 75.6&55550 & 0.076 &0.994  \\
\hline 68.7&55407 & 0.088 & 0.996 \\
\hline 61.1&55339 & 0.018 &0.997  \\
\hline 52.4&54689 & 0.059 & 0.973 \\
\hline
\end{tabular}
\caption{Photons from the vicinity of 4C +55.17 above 50 GeV observed by the \textit{Fermi} LAT instrument over the four year mission.  These photons are noted along with the probability for association with the FSRQ.}
\label{l2ea4-t1}
\end{center}
\end{table}

\section{VERITAS Observations of 4C +55.17}
VERITAS observed the FSRQ 4C\,+55.17 for 45 hours of livetime between May 2010 and March 2012 at an average zenith angle of 27 degrees, with observations optimized for a low energy threshold of 150 GeV.  These observations result in no VHE gamma-ray signal (0.0$\sigma$). A 95\% confidence upper limit is derived using the Rolke method \citep{8} for an assumed spectral photon index. Although the VERITAS limit is dependent on the assumed index of five, the limit changes by less than 10\% if derived for an index of $\Gamma$=5$\pm$0.5.  

We show the VHE flux from 4C\,+55.17 to be less than 2.3$\times10^{-12}$ ergs cm$^{-2}$ s$^{-1}$ at the decorrelation energy of 200 GeV during the VERITAS exposures, which is consistent with the \textit{Fermi} LAT extrapolations with no additional intrinsic turnover
above 1 GeV.  Such intrinsic turnover might expected from gamma-ray absorption by the FSRQ broad line region, as described in \cite{9} and \cite{10}.  It can be seen in Figure 2 that the VERITAS exposure is not yet deep enough to significantly constrain the VHE spectrum of the FSRQ. 

\begin{figure*}[t]
\centering
\includegraphics[width=160mm]{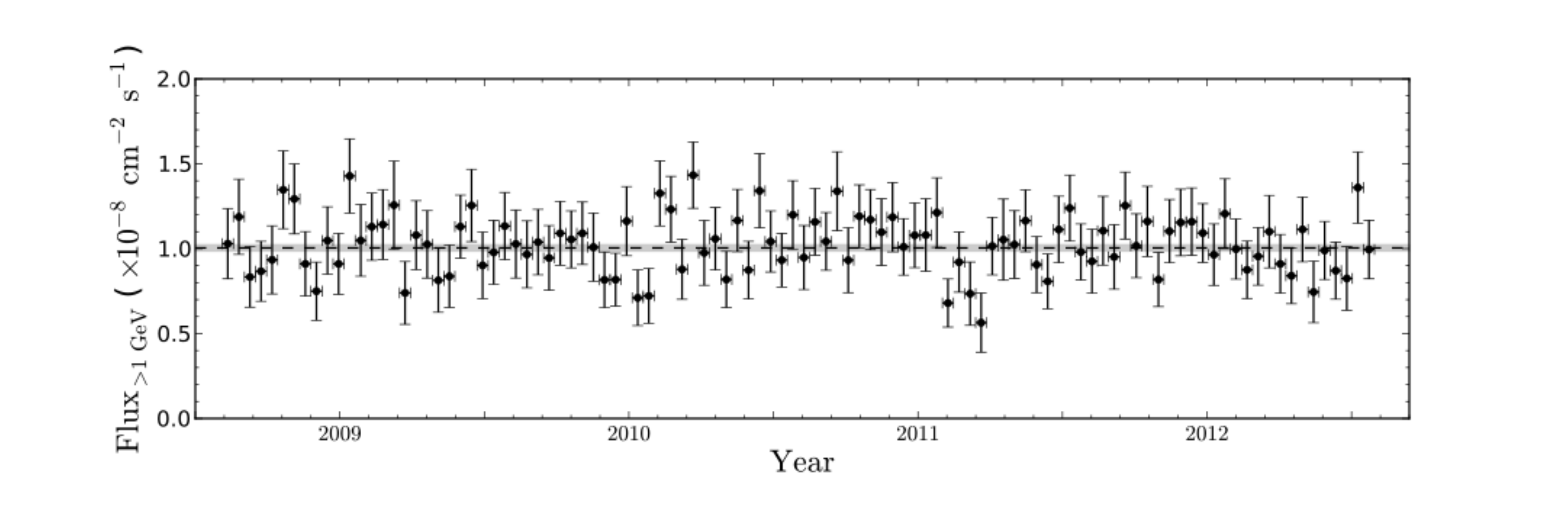}
\caption{A \textit{Fermi} LAT two-week bin
light curve of 4C\,+55.17 above 1 GeV. The
source shows no variability with a $\chi^2$ of 89
with 103 degrees of freedom, consistent at 83\% confidence level with steady state emission.} \label{fig1}
\end{figure*}

\begin{figure*}[t]
\centering
\includegraphics[width=150mm]{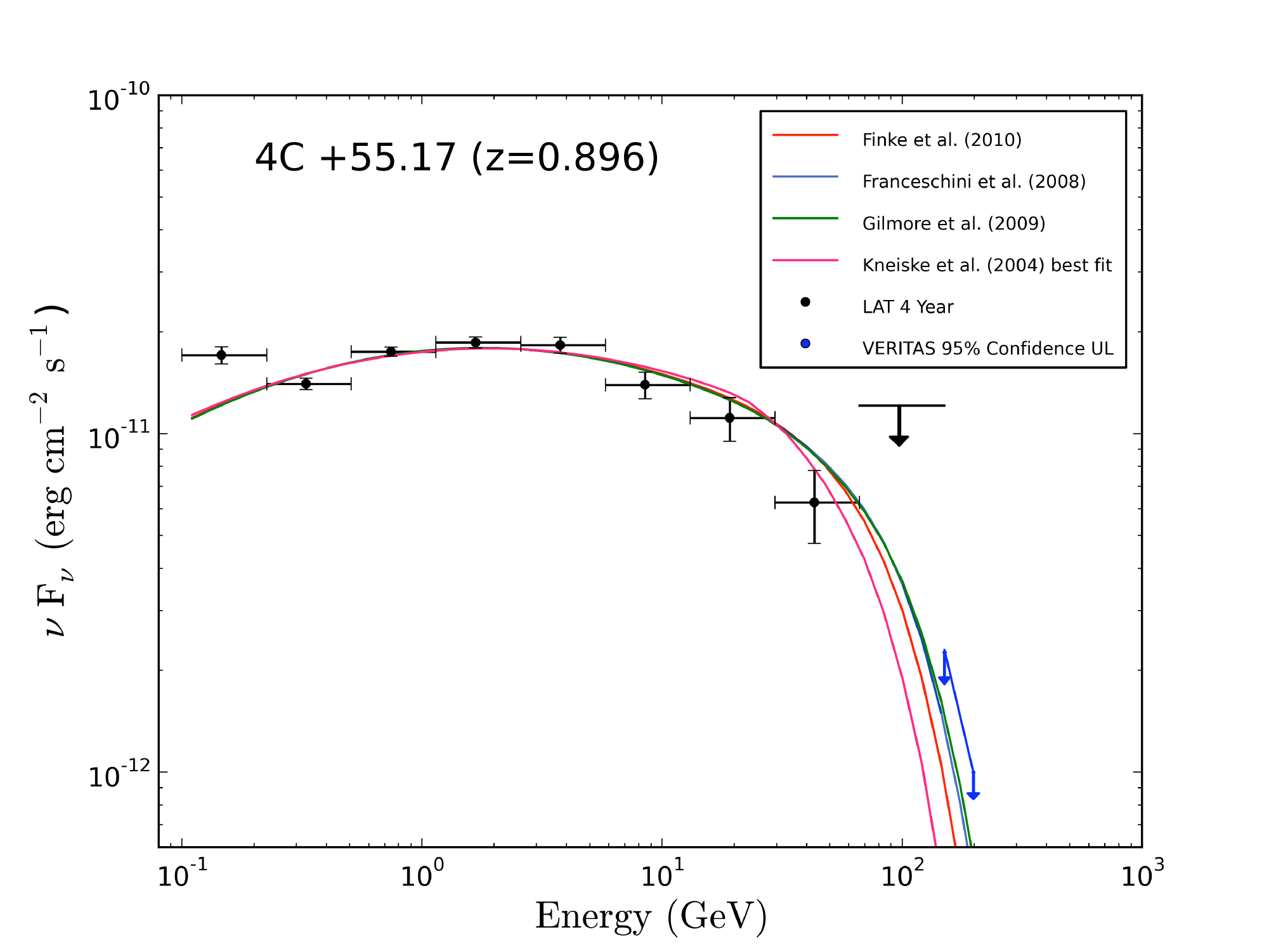}
\caption{The
spectrum and log-parabolic fit of 4C \,+55.17 above 100 MeV
including four years of LAT data. This spectrum is shown with the VERITAS 95\% confidence upper limit above 150 GeV,
derived assuming an index of $\Gamma$=5 for the differential power law $dN/dE=(E/E_{o})^{-\Gamma}$.} \label{fig2}
\end{figure*}

\bigskip 
\begin{acknowledgments}
This research was supported in part by NASA grant NNX10AP71G from the Fermi Guest Investigator program.  The VERITAS collaboration is supported by
grants from the US Department of Energy Office of
Science, the US National Science Foundation, and the
Smithsonian Institution, by NSERC in Canada, by
Science Foundation Ireland, and by STFC in the UK.
We acknowledge the excellent work of the technical
support staff at the FLWO and at the collaborating
institutions for the construction and operation of the
instrument. The Fermi Collaboration acknowledges
generous support from a number of agencies and
institutes that have supported the development and
the operation and scientific data analysis.
\end{acknowledgments}

\bigskip 

\end{document}